\documentclass[aps,prl,twocolumn,groupedaddress]{revtex4}
\usepackage[dvips]{epsfig}
\begin{document}

\title{Ordering in Heisenberg Spin Glasses}
\author{Doroth\'ee Petit, L. Fruchter}
\affiliation{Laboratoire de Physique des Solides,
\\ Universit\'e Paris Sud, 91405 Orsay, France\\}
\author{I.A. Campbell}
\affiliation{Laboratoire des Verres, \\ Universit\'e Montpellier II,
       34095 Montpellier, France}

\date{\today}

\begin{abstract}

For five different Heisenberg spin glass systems, torque experiments were performed in applied magnetic fields up to $4 T$. The Dzyaloshinski-Moriya random anisotropy strengths, the in-field torque onset temperatures, and the torque relaxation were measured. Critical exponents were estimated independently using a standard protocol. The data are strong evidence for a true spin glass ordered state which survives under high applied magnetic
fields; they can be interpreted consistently in terms of a chiral ordering model with replica symmetry breaking as proposed by Kawamura and coworkers.

\end{abstract}


\maketitle

The earliest experimental examples of spin glass (SG) ordering
were dilute magnetic
alloys with Heisenberg local spins; however
recently much more work has been carried out on Ising systems,
particularly through sophisticated numerical methods. For Ising
spin glasses (ISGs) in dimension three, large scale numerical
measurements show unequivocally that in zero field there exists a true critical
transition at a finite temperature $T_{g}$
\cite{Og,KY,pala,pom,bal}. For some time now theoretical
attention has been focused on a longstanding controversy as to
the correct description of the physics of the ISG
frozen state : does it resemble the Replica Broken Symmetry (RSB)
solution of the mean field model \cite{parisi}, has it a much
simpler structure described by the scaling or "droplet" approach
\cite{FH}, or is the correct description intermediate \cite{KMPY}
? In-field properties should distinguish between RSB and scaling
models, because for RSB there should be a phase transition
line under an applied magnetic field, while for the scaling model the
true phase transition exists only in zero field, and any apparent
transition line under field is a transitory relaxation effect. However because of numerical difficulties, there is still no definitive consensus for 3d ISGs \cite{marinari1,jerome,krz}.

For a \textit{Heisenberg} SG (HSG) in 3d, numerical
work appears to demonstrate conclusively that there should be no
ordering of the pure Edwards-Anderson type until zero temperature
\cite{olive,matsubara1,kawamura1,kawamura2}. Experiments
on the other hand clearly demonstrate that order at finite
temperatures exists in these systems, at least under zero
applied field. By exploiting the magnetic torque technique on
HSGs with a range of random anisotropy
strengths, we show that the SG order existing in zero
field survives under strong applied magnetic fields.
We have also estimated critical 
exponents of the same set of samples. The experimental data are compared 
with the predictions of the chiral order mechanism
proposed by Kawamura and co-workers
\cite{kawamura1,kawamura2,kawamura3} for HSGs,
which is a one step RSB model.

We have carried out extensive torque measurements on samples of
the textbook alloy SGs \textbf{Ag}Mn, \textbf{Cu}Mn,
\textbf{Au}Fe,  on a thiospinel insulating SG $CdCr_{2-x}In_{x}S_4$, and on an amorphous alloy SG
FeNi(PBAl). While there
have been numerous longitudinal magnetisation and susceptibility
measurements with increasingly subtle protocols on many SGs, torque experiments (which are essentially equivalent to
transverse magnetization measurements) can provide complementary
information on transverse magnetization irreversibility and
relaxation \cite{NdeC}, inaccessible to techniques where the field
orientation is held fixed. In a polycrystalline HSG the torque signal only exists because of the local random
Dzyaloshinski-Moriya (DM) interaction \cite{fertlevy}. The DM
anisotropy is due to a sum of spin-spin terms of the form
$\textbf{D}_{ij}\cdot(\textbf{S}_{i} \times \textbf{S}_{j})$ where
$\textbf{D}_{ij}$ represents a spin orbit interaction on a third
site which can be magnetic or not. For a given instantaneous set
of spin orientations $\textbf{S}_{i}(t)$ there will be
conventional spin-spin interaction terms plus the DM terms, and
the system on cooling will try to organise itself to minimise the
total energy. Suppose the sample is cooled in an applied field
$\textbf{H}_{0}$ to a temperature low enough for the spin system
to become rigid, so each local spin has taken up a fixed
orientation and all the spins are clamped together. If the field
is now turned through a small angle, the rigid spin system will
rotate bodily (the relative orientations of the spins will not
change) so that the remnant magnetisation $\textbf{M}_{r}$ turns
towards the new direction $\textbf{H}$ of the field and takes up
an orientation which minimises the total energy $\textbf{H} \cdot
\textbf{M}_{r}$ plus the sum of the rotated DM terms, and which
is not parallel to $\textbf{H}$. With no internal spin
reorganisation  there will just be an effective unidirectional
anisotropy $K$ acting on the spin system \cite{fertlevy}. In
other words there will now be a transverse component of the
magnetisation perpendicular to the new direction of the field
which gives rise to a torque signal 
\begin{equation}
\textbf{$\Gamma$}=\textbf{H}\times\textbf{M}
\end{equation}
The intrinsic anisotropy of a sample can be estimated
by extrapolating the torque measurements to zero temperature
where the rigid spin system torque will saturate with field at a
field and time independent value $K(0)\theta$. At non-zero T and a frozen-in spin system there will be 
a quasi-static torque after the turning
procedure with magnetic creep of the type familiar from residual
magnetisation measurements in SGs. When the spin system is not
rigid but can reorganise during the rotation of the field because
of rapid local spin relaxation, no torque will be observed. A field-temperature phase
diagram can be drawn using 
the observation of non-zero torque on a chosen time and sensitivity scale as a criterion for a phase boundary.

We  use a specially constructed torquemeter with a symmetric
axial suspension, a horizontal superconducting Helmholtz coil to
provide the field, and a capacity bridge detector. The principal
protocol used was to field cool (FC) the sample from well above
$T_{g}$ in a fixed orientation applied field down to the measuring
temperature $T$. After the sample temperature reached
equilibrium, the field was turned typically through $5^{o}$ in $20$
seconds. The torque signal was monitored for the range of
times from 10 seconds to 3600 seconds after the turn was completed. 
 The zero field
transition temperatures $T_{g}$ of the samples and the normalized $K(0)$ are given in Table 1.
\begin{table}[htbp] 
\caption{\label{Table:1} Values of the freezing temperatures $T_g$, and the ratio of anisotropy constant (in erg/mol of magnetic sites) to $T_g$.} 
\begin{ruledtabular}
\begin{tabular}{ccc}
sample & $T_g$(K)
&$K(0)/T_g \times 10^{-5}$ \\ 
\textbf{Cu}Mn $3\%$ &18.5 &0.068 \\ 
\textbf{Ag}Mn $3\%$ & 11.9   
&0.16  \\ 
CdCr$_2$InS$_4$  &16.8  &0.8 \\ 
\textbf{Au}Fe $8\%$  &23.9 &
1.32 \\ 
(Fe$_{0.1}$Ni$_{0.9}$)$_{75}$P$_{16}$B$_6$Al$_3$ &13.4 &2.65 \\ 
\end{tabular}
\end{ruledtabular}
\end{table}

\begin{figure}[h]
\includegraphics[height=6cm,width=9cm,angle=0]{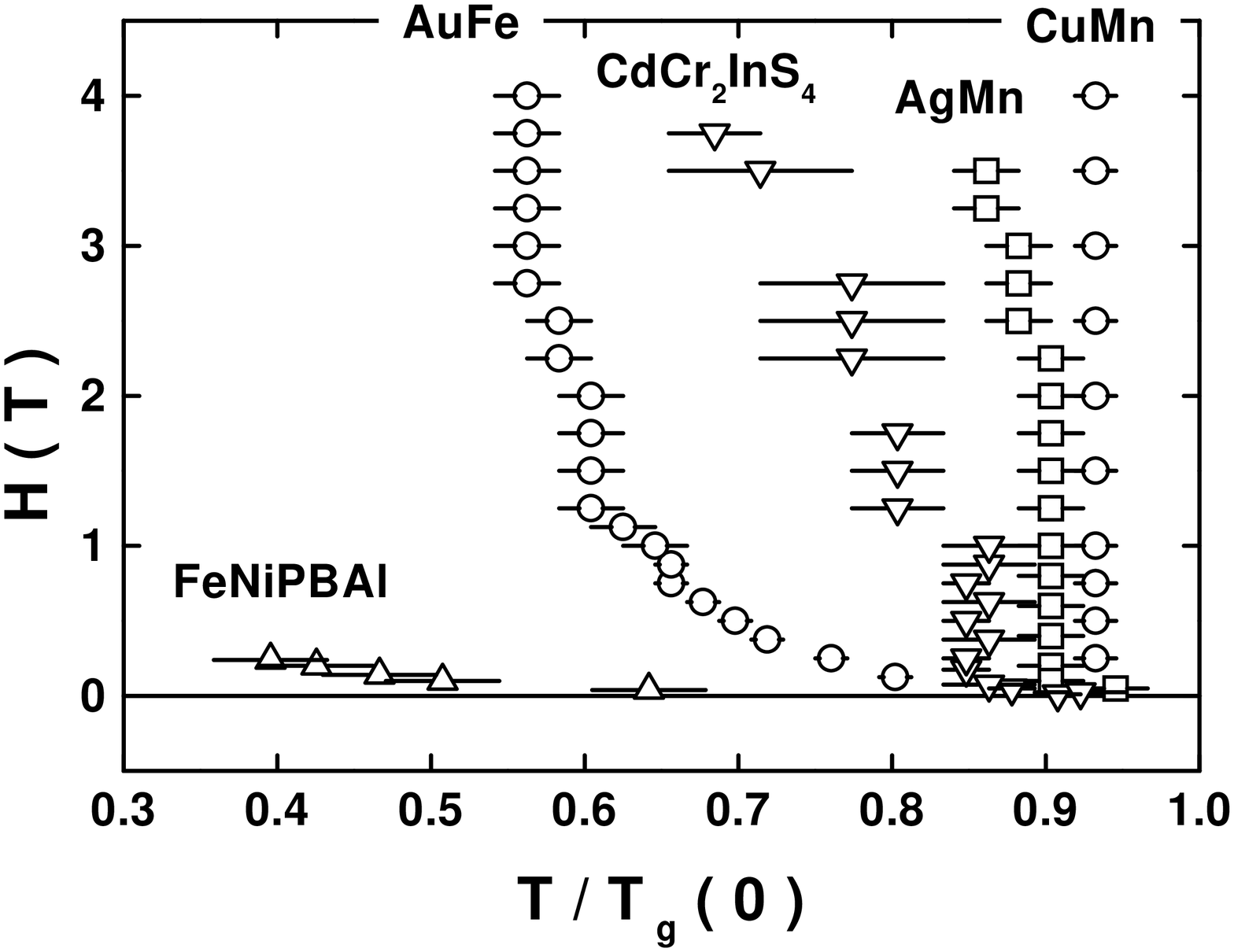}
\caption{The normalized onset temperatures for torque as a function of field for the different samples \label{Figure:1}} 
\end{figure}

We show in Figure 1 normalised effective phase lines for the
different samples. To define the transverse
irreversibility onset for this figure the torque measuring time was chosen as $100$
seconds after turning and the onset was defined to occur when the signal dropped to less than $10^{-3}K(0)$. The "error bars" in Figure 1 correspond to a plausible range of onset criteria are used. We have also estimated the longitudinal irreversibility
onset for the same samples using standard FC and ZFC measurements
performed with a SQUID magnetometer. The magnetisation
irreversibility results are consistent with many other
observations of this type in the literature, in particular the "weak
irreversibility" line of \cite{kenning}. There is a large section
of the field-temperature plane where transverse irreversibility
appears very clearly in the torque measurements, but where there
is no visible longitudinal irreversibility.
Comparing Table I and 
Figure 1 it is clear that
there is a systematic change in the form of the transverse onset lines with
anisotropy. Transverse and longitudinal (not shown in the figure)  irreversibility lines
fuse at progressively lower $T/T_g$ as the anisotropy increases. For the systems which are in the weak
anisotropy limit, a quasi-vertical section of the transverse onset
line sets in at a temperature marginally below $T_g$. Data on the moderately strong anisotropy sample {\textbf Au}Fe were discussed in \cite{dorothee}.
The relaxation of the torque signal was studied systematically for all samples. As an example of the behaviour seen in a weak anisotropy system,  \textbf{Ag}Mn torque relaxation data taken at $H = 3 T$
for temperatures $T$ ranging from $4.5 K$ to $9.5 K$ are shown in Figure 2.
\begin{figure}[h]
\includegraphics[height=6cm,width=9cm,angle=0]{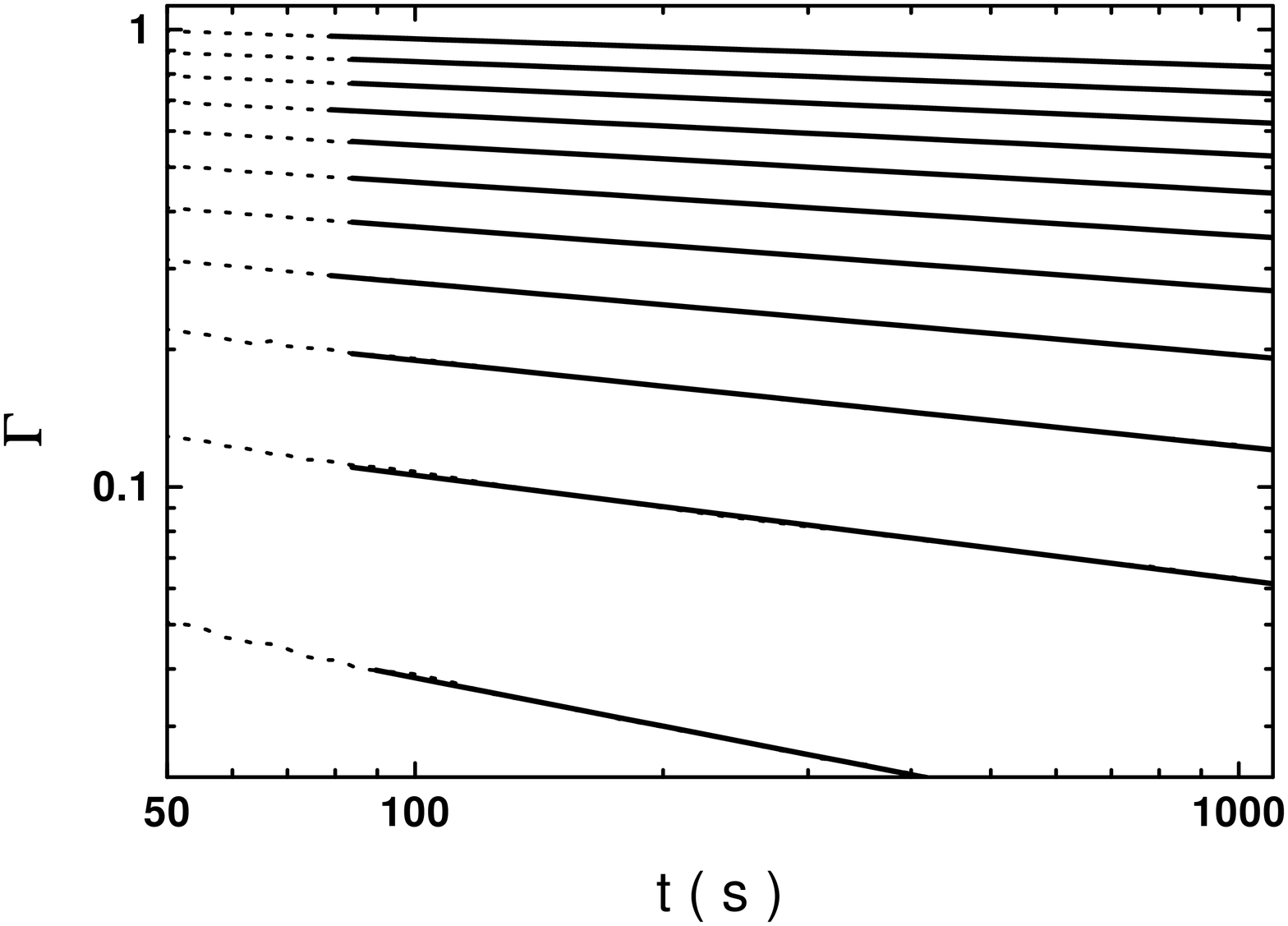}
\caption{The relaxation of the torque signal at $H=3T$ for the \textbf{Ag}Mn sample on a log-log plot. Temperatures from top to bottom go from $4.5 K$ to $9.5 K$ by steps of $0.5 K$\label{Figure:2}} 
\end{figure}
 It can
be seen that in all cases the relaxation is essentially pure
algebraic, 
$\Gamma(t) \sim t^{-\alpha}$, so no relaxation time can be defined. $\alpha$ increases slowly with $T$, so the relaxation ressembles the
well known zero field remanent magnetisation relaxation.
At each temperature $\alpha$ is almost field
independent, Figure 3. No aging effects have been observed 
for torque relaxation after FC. 
These relaxation data are in striking contrast to magnetisation
relaxation results reported for an ISG sample under
applied field \cite{mattsson}. In the Ising case an effective relaxation
time $\tau(H)$ was extracted from ac susceptibility measurements and was found
to drop exponentially with increasing applied field. For instance at the temperature $\frac{T}{T_{g}} = 0.7$, at $H = 0.3 T$
$\tau(H)$ was 300 seconds while by $H = 1.5 T$, $\tau(H)$ had dropped to
about $3$ milliseconds;  an extrapolation to $H = 4T$ would lead to
something like $\tau(H) \sim 10^{-10}$ seconds. The Ising
behaviour was interpreted in terms of the droplet approach as
showing that there was no real transition in field but only a
pseudo-transition which could be put down to strongly field dependent relaxation effects.
\begin{figure}[h]
\includegraphics[height=6cm,width=9cm,angle=0]{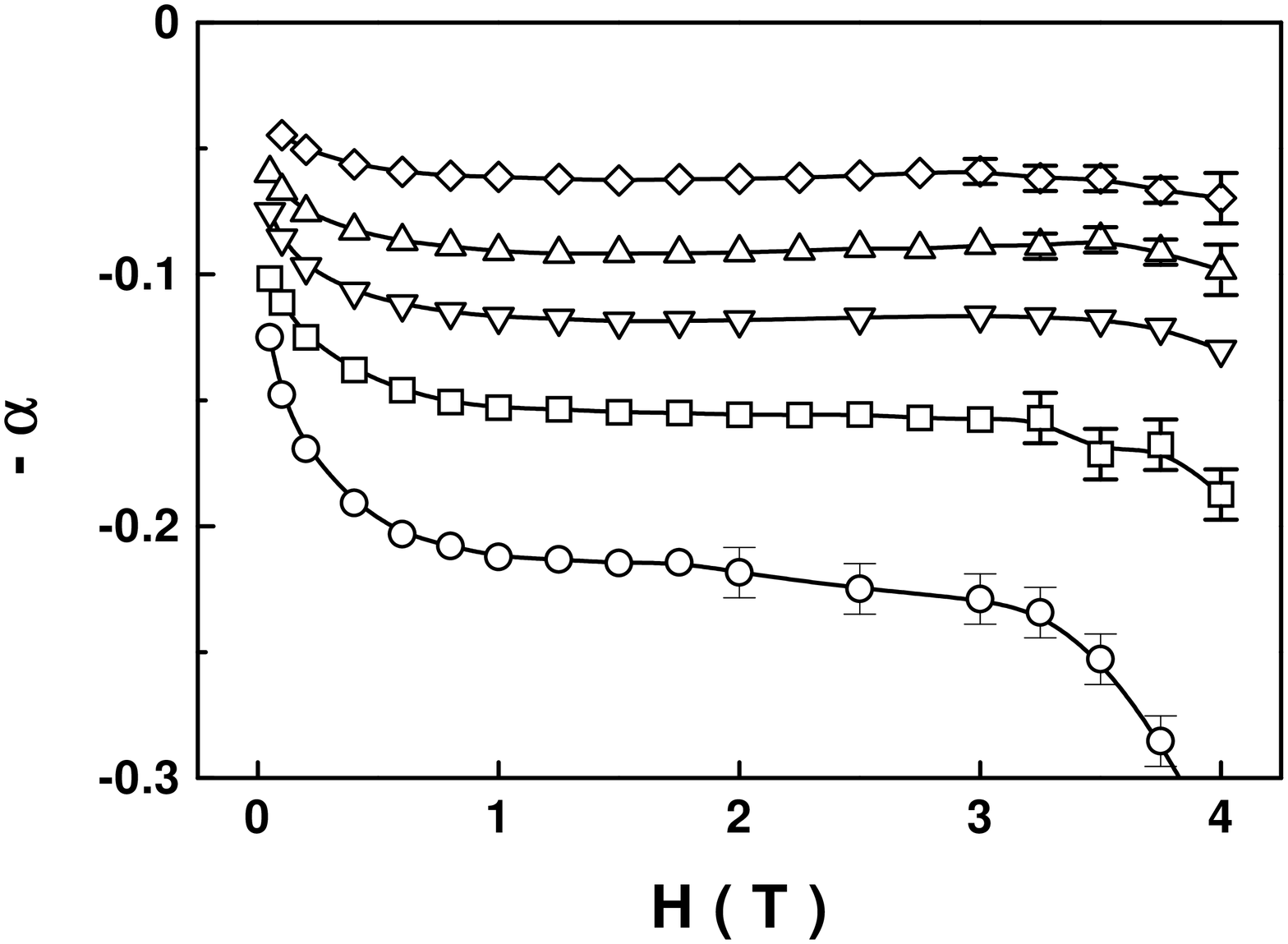}
\caption{The algebraic relaxation exponent $\alpha$ as a function of field for temperatures which are from top to bottom 4.5,6,7,8,and 9 K.\label{Figure:3}}
\end{figure}
\textit{A contrario} in the present weak anisotropy HSG
case, for a wide range of fields the torque onset is field independent, the
torque relaxation has no time scale and $\alpha$ is nearly field independent. This behaviour can be taken as characteristic of a true in-field phase transition.\\
It has already been noted that published experimental critical exponent estimates for spin glasses show wide variations. 
We have carried out static magnetisation measurements on our five samples using a commercial SQUID magnetometer in order to estimate static critical exponents. The protocol used was identical in all cases and followed that established by \cite{bouchiat}. There are no time dependencies in these measurements. In each case where comparisons can be made the exponent results from the present work are in excellent agreement with the published estimates \cite{levy,bouchiat,hammann,gunnarsson}. All measurements indicate strongly positive $\eta$ values for the weak anisotropy cases.
The details of the analysis of the present data will be presented elsewhere \cite{dorothee2}. 

We will discuss the results in terms of the 
chiral ordering model of Kawamura and coworkers \cite{kawamura1,kawamura2,kawamura3}. Chiral ordering was invoked as a possible mechanism for HSGs some time ago \cite{olive} ; the influence of weak DM interactions on SG ordering has also been studied in the context of dilute magnetic semiconductors \cite{marinari2,larson}; see also \cite{matsubara2}. 
In the Kawamura model the mechanism of HSG ordering is through the chiral order parameter. The numerical data indicate that the form
of the ordering is one step RSB, and that the pure chiral ordering is very
robust against applied magnetic fields \cite{kawamura3}. A pure
chiral ordering would be invisible experimentally but it would
be  "revealed" by even a weak local anisotropy term such as DM anisotropy
which links together the chiral order parameter and the spin
order parameter. As a rule of thumb, the weaker the anisotropy the 
more similar the qualitative behaviour should appear to that of the pure chiral case though detailed numerical calculations as a function of the DM interaction strength and applied field have not not been performed so far.  
The field-temperature phase diagram should
mimic that of the HSG mean field model \cite{MF} with distinct quasi
"Almeida-Thouless" longitudinal and "Gabay-Toulouse" transverse
irreversibility lines (though the physical mechanisms in the mean
field and chiral models are quite different \cite{kawamura3}). As in the mean field model, the transverse irreversibility onset represents a
true phase transition. Also as in the mean field model, for low
fields such that the anisotropy dominates (i.e. $HS < DS^2$) ,
the longitudinal and transverse irreversibility lines  should
fuse. This fusing should thus extend from very near $T_g$ for low anisotropy down to much smaller values of
$\frac{T}{T_{g}}$ for systems where the anisotropy is stronger.
For strong applied fields the transverse irreversibility line
rises well above the longitudinal irreversibility line, finally
saturating at a high field for low temperatures \cite
{kawamura3}. The qualitative agreement between model and experiment is striking; both indicate that the weak anisotropy SG
ordering is very robust against applied magnetic fields. The model also predicts the observed relative weakening of the transverse irreversibility  with increasing anisotropy. A satisfactory quantitative comparison can be made between the pure chiral 
model and the weak anisotropy sample torque experiments \cite{dorothee2}. It can also be noted that the numerical work on the pure chiral model \cite{kawamura2} indicates a strongly positive critical exponent $\eta$, similar to the strongly positive $\eta$ values obtained from the analysis of the experimental data on the weak anisotropy systems.  

Turning to the strong anisotropy cases such as the $FeNi(PBAl)$ sample, ordering is much less robust under an applied field, as for the $FeMnTiO_3$ ISG system \cite{mattsson}. Numerical work also indicates that if there is true in-field ordering in a 3d ISG it is rapidly destroyed by a magnetic field \cite{krz}. In view of the similar low field memory properties observed in HSGs and ISGs \cite{vincent}, it would be
surprising if the physics of the two types of system were
fundamentally different (RSB for the HSGs, "droplet" for the ISGs). 

In conclusion, we have made detailed measurements of the DM anisotropy, the  transverse irreversibility onset and relaxation, and the critical exponents for five different HSGs. 
We find that the behaviour is strongly anisotropy dependent; for weak anisotropy the tranverse irreversibility onset is very robust under strong applied magnetic fields, which is the sign of in-field spin glass ordering characteristic of RSB whatever the detailed physical model. 
It is perhaps the first time that unequivocal evidence can be given in
favour of RSB ordering in the experimental SG context. Although further numerical work involving anistropy and applied fields would be most welcome, a satisfactory solution to the 
HSG ordering enigma appears to be the chiral 
model \cite{kawamura1,kawamura2,kawamura3} which is in the RSB class.

\end{document}